# The High Energy X-ray telescope (HE) onboard the *Insight*-HXMT astronomy satellite

CongZhan Liu [1*], YiFei Zhang [1††], XuFang Li [1††], XueFeng Lu [1††], Zhi Chang [1††], ZhengWei Li [1††], Aimei Zhang [1††], YongJie Jin [2††], HuiMing Yu [2††], Zhao Zhang [2††], MinXue Fu [2††], YiBao Chen [2††], JianFeng Ji [2††], YuPeng Xu [1,3††], Jingkang Deng [2], RenCheng Shang [2], Guoqing Liu [2], FangJun Lu [1], ShuangNan Zhang [1,3], YongWei Dong [1], TiPei Li [1,2,3], Mei Wu [1], YanGuo Li [1], HuanYu Wang [1], BoBing Wu [1], YongJie Zhang [1], Zhi Zhang [2], ShaoLin Xiong [1], Yuan Liu [1], Shu Zhang [1], HongWei LIU [1], YiRong Yang [1], Fan Zhang [1]

[1] *Key Laboratory for Particle Astrophysics, Institute of high energy physics, Chinese Academy of Sciences, Beijing 100049, China;*
[2] *Tsinghua University, Beijing 100084, China;*
[3] *University of Chinese Academy of Sciences, Chinese Academy of Sciences, Beijing 100049, China*

The *Insight*-Hard X-ray Modulation Telescope (*Insight*-HXMT) is a broad band X-ray and gamma-ray (1-3000 keV) astronomy satellite. The High Energy X-ray telescope (HE) is one of its three main telescopes. The main detector plane of HE is composed of 18 NaI(Tl)/CsI(Na) phoswich detectors, where NaI(Tl) serves as primary detector to measure ~ 20-250 keV photons incident from the field of view (FOV) defined by the collimators, and CsI(Na) is used as an active shield detector to NaI(Tl) by pulse shape discrimination. CsI(Na) is also used as an omnidirectional gamma-ray monitor. The HE collimators have a diverse FOV: 1.1°x 5.7° (15 units), 5.7°x 5.7° (2 units) and blocked (1 unit), thus the combined FOV of HE is about 5.7°x 5.7°. Each HE detector has a diameter of 190 mm, resulting in the total geometrical area of about 5100 cm$^2$. The energy resolution is ~15% at 60 keV. The timing accuracy is better than 10 μs and dead-time for each detector is less than 10 μs. HE is devoted to observe the spectra and temporal variability of X-ray sources in the 20-250 keV band either by pointing observations for known sources or scanning observations to unveil new sources, and to monitor the gamma-ray sky in 0.2-3 MeV. This paper presents the design and performance of the HE instruments. Results of the on-ground calibration experiments are also reported.

# 1 Introduction

The Hard X-ray Modulation Telescope (HXMT), dubbed *Insight*-HXMT after being launched on June 15, 2017, is China's first X-ray astronomy satellite devoted to broad band observations in the 1-3000 keV band. In order to fulfill the requirements of the broad band spectral and temporal variability observations of the X-ray and gamma-ray sky, *Insight*-HXMT carries three main telescopes: High Energy X-ray telescope (HE) using an array of NaI(Tl)/CsI(Na) scintillation detectors, covering the 20-250 keV energy band, for pointing and scanning observations, with an extension up to ~3 MeV for gamma-ray monitoring, Medium Energy X-ray telescope (ME) using an array of Silicon Positive-Intrinsic-Negative (Si-PIN) detectors in the 5-30 keV band, and Low Energy X-ray detector (LE) using an array of Swept Charge Device (SCD) detectors in the 1-15 keV band [1-3]. These three telescopes are co-aligned in the same direction to simultaneously observe the same sources, although the FOV of each telescope is slightly different. To perform cross-calibration and measure the shape of broad band spectra more accurately, there is some overlap between the energy bands of HE and ME, and also for ME and LE.

Since the 1970s, several tens of high energy astronomical satellites have performed comprehensive studies in the X/γ-ray band and achieved highly significant progresses. The important satellites or instruments in the hard X-ray band include HEAO1/A4, Mir/HEXE, CGRO/BATSE, CGRO/OSSE, RXTE/HEXTE, BeppoSAX/PDS, Suzaku/HXD, INTEGRAL/IBIS, Swift/BAT, and NuSTAR [4-7]. The early instruments focused on spectral and temporal variability observations. In recent years, INTEGRAL and Swift have performed all-sky X/γ-ray sources surveys; NuSTAR has achieved a breakthrough in hard X-ray focusing technology and obtained high resolution images. These satellites have investigated these compact objects in great detail [4, 7].

Similar to BeppoSAX/PDS and RXTE/HEXTE, *Insight*-HXMT/HE adopts an array of NaI(Tl)/CsI(Na) phoswich detectors as the main detector plane, with a total geometric area of about 5100 cm$^2$ and a combined Field of View (FOV) of about 5.7°×5.7° (FWHM). The CsI(Na) in the phoswich of HE can be also used as a γ-ray burst (GRB) monitor. The measured deposition energy range of CsI(Na) is about 40-600 keV in the normal mode and about 200 keV - 3 MeV (deposited energy in this paper) in low gain mode (previously called GRB mode). It is expected to play an important role in the observations of GRBs and electromagnetic counterparts of Gravitational Wave (GW) events found by LIGO and Virgo [8].

In this paper, we present the design of the HE instrument, its basic performances, and the preliminary results of on-ground calibration experiments.

# 2 Instrument description

The HE telescope consists of the main detector (HED), high energy collimator (HEC), auto-gain control detector (HGC), anti-coincidence shield detector (HVT), particle monitor (HPM), data processing and control box (HEB) and power box (HEA). The view of the HE experiment aboard *Insight*-HXMT is shown in Figure 1 and main characteristics of the HE are listed in Table 1.

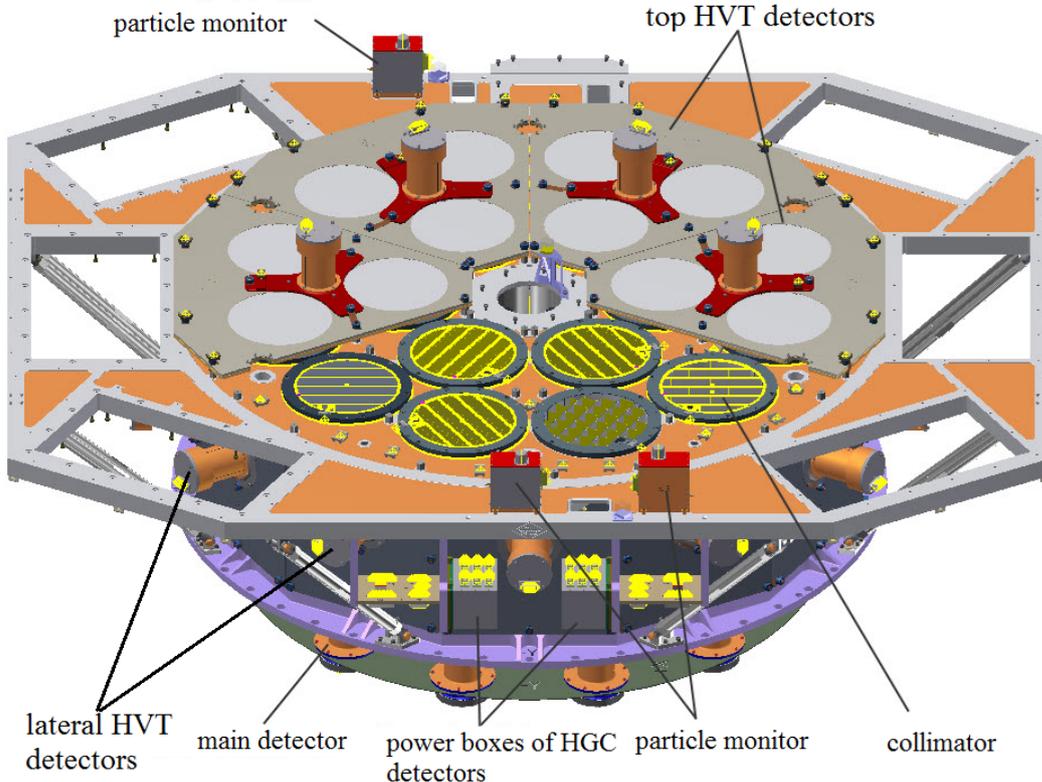

Figure 1. Illustration of the HE telescope onboard the *Insight*-HXMT astronomy satellite. The HE telescope is composed of main detectors, collimators, auto-gain control detectors (not shown), anti-coincidence shield detectors (HVT) and particle monitors.

Table 1 Main characteristics of HE

| Parameter | Values |
| --- | --- |
| Energy band | 20-250 keV (normal mode)<br>0.2-3 MeV (low gain mode) |
| Geometric area | 5096 cm$^2$ |
| Main detector | NaI(Tl)/CsI(Na)<br>~ 3.5 mm/40 mm |
| Timing accuracy | < 10 μs |
| Deadtime | < 10 μs |
| Combined FOV (FWHM) | 5.7°x 5.7° |
| Energy resolution (FWHM) | ~15%@60 keV |
| Maximum count rate | >30,000 cnts/sec |

As the main detector of HE, HED is responsible for the observation of celestial sources. Eighteen collimators are arranged in a pattern shown in Figure 2. HGC provides the auto-gain control and energy calibration for HED. HVTs act as an active shielding system to reduce HED's background caused by charged particles. HPMs monitor the flux of charged particles and send out an alert to switch off the high voltages of HEDs and HVTs in high flux regions to avoid potential damage to the PMTs of these detectors.

The signals from all HE detectors are dealt with by the HEB. The data are delivered to the satellite platform by Low Voltage Differential Signal (LVDS) and 1553B bus. HEB receives the time-synchronous command, pulse-per-second (PPS) signal of GPS, and a high-accuracy 5 MHz clock signal from the satellite platform. HEA provides power for all detectors and HEB.

**2.1 The High Energy Detector (HED)**

The HED adopts an NaI(Tl)/CsI(Na) phoswich as the detector. The phoswich technique was invented in the early 1950s and was employed by HEAO1/A4 telescope for the first space application [9, 10]. The scintillation detector NaI(Tl)/CsI(Na) phoswich is a mature and widely used technology in high energy astrophysics, e.g., HEAO1/A4, Mir/HEXE, BeppoSAX/PDS and RXTE/HEXTE [5-7,10]. The light yield of NaI(Tl) is fairly high and the detection efficiency is also high for hard X-rays. Its decay time of scintillation is 230 ns at room temperature. The peak of the emitting spectrum is around 410 nm. It is strongly hygroscopic and thus needs to be hermetically sealed. The light yield of CsI(Na) is about 85% of that of NaI(Tl). The decay time of CsI(Na) is 630 ns. The peak of emitting spectrum is around 420 nm [11]. Since the emitting spectrum peaks of these two types of crystals are very close, they are suitable for being coupled together to compose a phoswich detector. The scintillation signals of these two types of crystals can be read out by a shared Photomultiplier Tube (PMT) and then selected by Pulse Shape Discrimination (PSD) technique to pick up a "good event" in which energy is deposited only in the NaI(Tl) scintillator with no detectable signal in CsI(Na). For HE, CsI(Na) is primarily used as an active shielding detector to reject Compton events which deposit energies in both CsI(Na) and NaI(Tl) simultaneously, and background events incident from backside within ~2π solid angle.

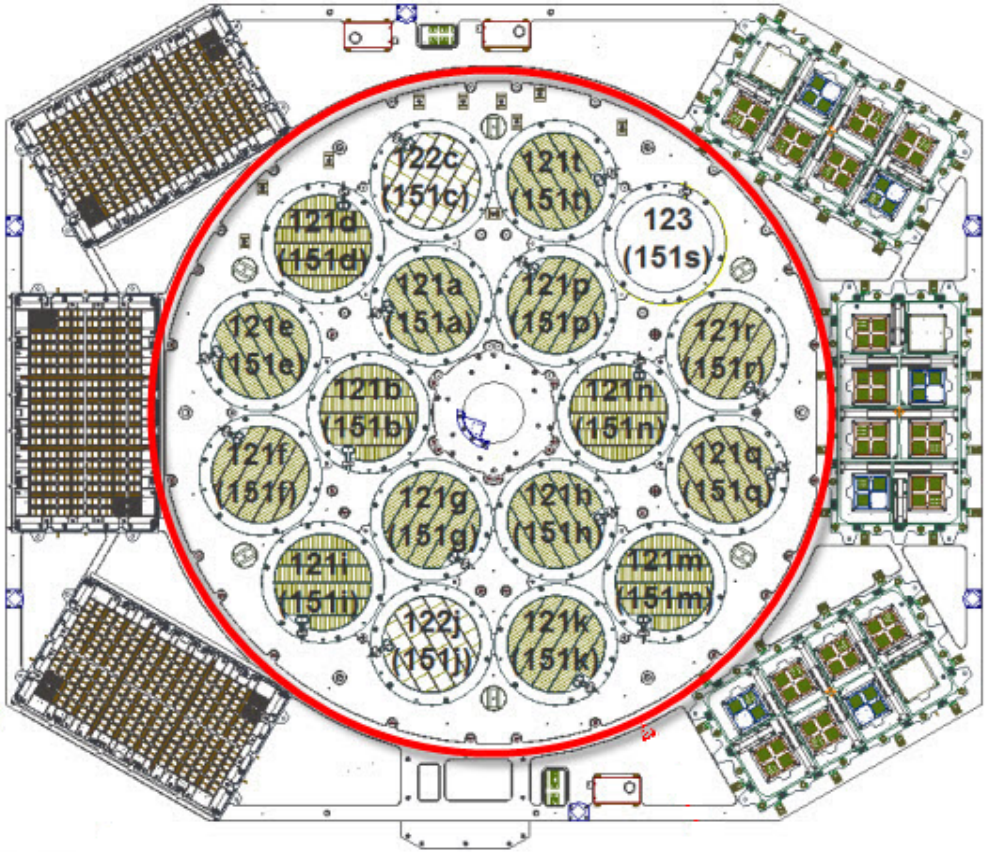

Figure 2. Distribution pattern of 18 HE collimators. The numbers beginning with "15" stand for collimators, and the numbers beginning with "12" stand for the corresponding auto gain control detectors. The three boxes on the left are ME telescope and the three boxes on the right are LE telescope.

The diameter of each scintillator in the HED is 190 mm and its geometric area is 283.5 cm$^2$. Thus, the total geometric area of 18 HEDs is about 5100 cm$^2$. The thickness of the NaI(Tl) is 3.5 mm, while the CsI(Na) is 40 mm. The whole phoswich is hermetically sealed in a box consisting of a beryllium window, aluminum alloy shell, and a 10 mm-thick quartz glass. As shown in Figure 3, the overall size of a single HED is 243 mm×243 mm×285 mm.

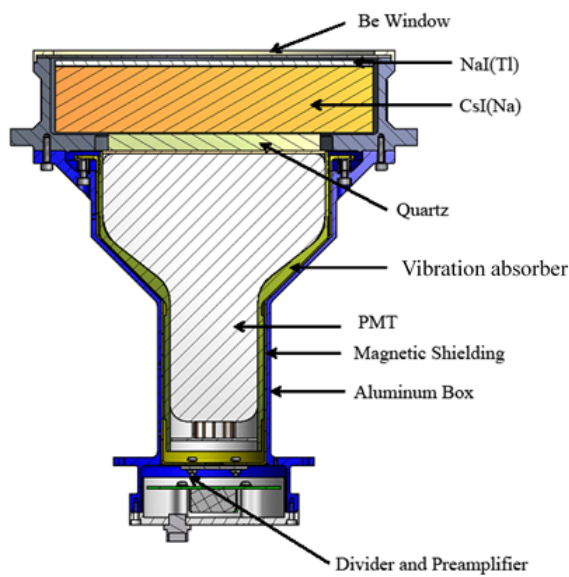

Figure 3. A cutaway view of an HED.

The entrance window is a 1.5 mm-thick beryllium plate, for which the transmissivity of 20 keV X-ray photons is higher than 90%. It can also satisfy the requirements of hermetical sealing and vibration. A PMT (Hamamatsu R877-01) is coupled with a quartz glass by a kind of silicone gel to read out the scintillation light. The PMT is surrounded by a vibration absorber and magnetic shielding to reduce the impacts of vibration and geomagnetic field. The PMT divider adopts a negative high voltage around -1000V. The signal is delivered into a preamplifier via DC coupling. The design of the divider and preamplifier are optimized to match the interface circuit of HEB and to improve the response and recovery time of HED to events with large deposited energy.

### 2.2 HE collimator (HEC)

HE is a collimating telescope designed to employ the direct demodulation imaging method [1]. The collimator needs a large FOV to achieve a high scanning efficiency (accumulate adequate X-ray photons in a short time); meanwhile the FOV cannot be too large to reduce the effect of modulation on the imaging resolution, and to increase the incident background.

There are three types of collimators according to their individual FOV. The FOV of 15 collimators is set to 1.1°x 5.7°(FWHM), the so-called "narrow-FOV collimators"; two collimators are set to 5.7°x 5.7°, the so-called "wide-FOV collimators". Specifically, one collimator with an FOV of 1.1°x 5.7° is covered by a 2 mm-thick tantalum lid, which is used to measure the local background of HE and is called a "blocked collimator". Eighteen collimators together with the 18 HEDs are distributed along two concentric circles (6 in the inner circle and 12 in the outer circle). The layout of the collimators in the supporting structure is shown in Figure 2. The collimators numbered by 122c and 122j are wide-FOV collimators; the collimator numbered by 123 is the blocked collimator. All the 18 collimators are divided into three groups: group A consists of the collimators numbered by 121a, 121b, 122c, 121d, 121e, and 121f (5 narrow-FOV collimators and 1 wide-FOV collimator); group B consists of the collimators numbered by 121g, 121h, 121i, 122j, 121k, and 121m (5 narrow-FOV collimators and 1 wide-FOV collimator); group C consists of the collimators numbered by 121n, 121p, 121q, 121r, 123, and 121t (5 narrow-FOV collimators and 1 blocked collimator).

The overall dimension of an individual collimator is Φ238 mm×300 mm; the outer frame of the HE collimator is fabricated with a monolithic aluminum alloy ingot. The top surface of the flange plate of this frame is considered as the reference plane to machine the cylinder and internal stiffeners, which ensures the mechanical precision and rigidity of the collimator. On the inner surface of the cylinder and stiffeners, slots for inserting 0.14 mm thick Ta (tantalum) sheets are machined by a wire-cutting technique; a 0.14 mm-thick Ta sheet is also fixed to these inner surfaces. According to simulations, the point spread function (PSF) of a collimator is slightly higher than the ideal value by 2.5% due to penetration of hard X-rays at 250 keV (less than 1% below 200 keV).

Figure 4 is a sketch of the three different kinds of collimators. There is a slot at the upper left corner of the aluminum alloy cylinder of each collimator, which is used to install the HGC and its cable; Figure 5 roughly shows the superposed visible area of the collimators for varying FOVs.

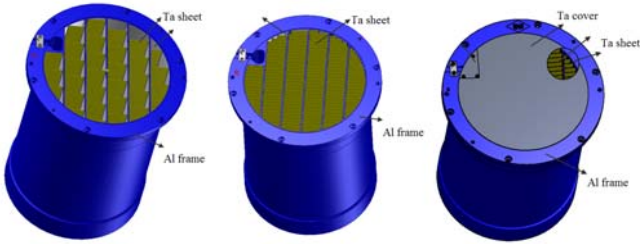

Figure 4. A structural sketch of the HE collimator, left: Wide-FOV collimator; middle: Narrow-FOV collimator; right: Blocked collimator

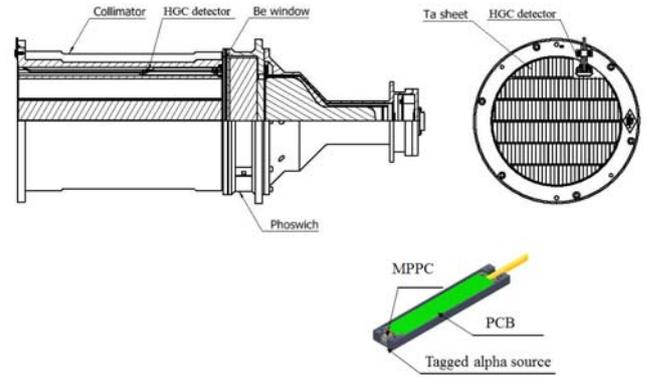

Figure 6. Configuration of HED, HEC and HGC. HGC is installed in a corner of the corresponding collimator. The bottom panel is a sketch map of an HGC.

Accompanying the 5.5 MeV α particle emitted during the decay of $^{241}$Am, a 59.5 keV X-ray photon will be simultaneously emitted according to the branching ratio [5, 11]. Thus, AGC can be accomplished by a coincident measurement with the 59.5 keV X-ray photon and the simultaneous α particle. The 5.5 MeV α particle loses its energy and produce fluorescence in the plastic scintillator, which is read out by an MPPC (Multi-Pixel Photon Counter, produced by Hamamatsu). At the same time, part of the accompanying photons with energy of 59.5 keV are detected by HED. Both the signals of HGC and HED are delivered into HEB. HEB will label it as a calibration event in the data packet if the two signals are coincident in a 2μs time window.

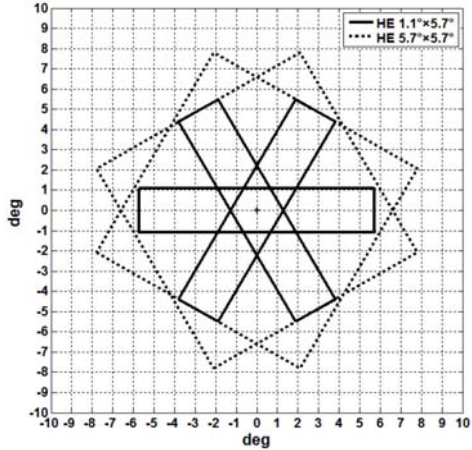

Figure 5. The superposed FOVs of the HE telescope. The three solid rectangles stand for the FOVs of the three sets; the two dashed squares represent the FOVs of the two wide-FOV collimators.

**2.3 Automatic gain control detector (HGC)**

There is an HGC in front of each HED. It plays the role of an Automatic Gain Control (AGC) for HED in orbit. A radioactive source $^{241}$Am (its half-life period is 433 years) with radioactivity of about 200 Bq is embedded into a small BC-448M plastic scintillator to act as a source of HGC. A whole HGC is installed in an aluminum shell, the overall dimension of which is 120×16×34 mm$^3$. Each HGC is fixed by three bolts in the corresponding collimator. Its size is small and will not significantly block the FOV of HED (see Figure 6).

Here is the principle of AGC. In normal mode of HE, the expected full-energy-peak generated in HED by calibration events is fixed at P0. A window is set centered at P0 with a width of 15 channels at each side as shown in Figure 7. If the ADC channel of a calibration event falls in the left window, the high voltage of HED will be increased by one step; if it falls into the right window, the high voltage of HED will be decreased by one step; if it falls out of the window, this event will be ignored and the high voltage of HED will not be changed. This will compensate the gain shift induced by external factors and achieve a long-term dynamic stability of gain.

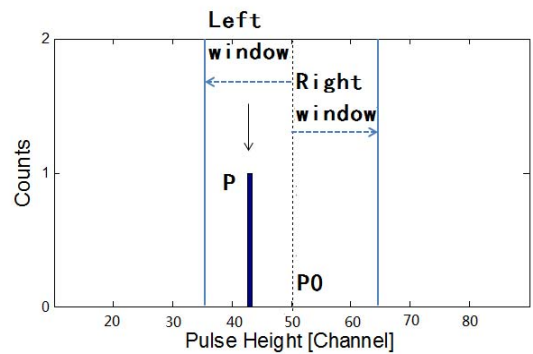

Figure 7. Schematic of AGC windows

The change of HED high voltage is 312.5 mV according to a step adjusted by HEB, which will roughly cause a ~0.3% correction of the PMT gain assuming the PMT voltage is -1000V. If the gain cannot be recovered after 128 successive steps in one direction, the system is considered to be abnormal and AGC will stop. This situation will be dealt with by uploading a command from the ground.

In fact, the HGC detector is much smaller than HED. The 59.5 keV X-ray photons can only inject a particularly small area of the HED phoswich. So, an issue is that the full-energy-peak of the calibration events will be inconsistent with that of the entire HED detector performance to the 59.5 keV photons. It is essential to understand the energy response nonuniformity of HED via an on-ground calibration test.

**2.4 Anti-coincidence detector (HVT)**

An HVT is basically made of a 6.4mm thick plate of plastic scintillator which is sensitive to the charged particles. There are eighteen HVTs covering the front and side faces of the HED to effectively shield the background induced by the incident charged particles from the front side $2\pi$ solid angle. To obtain sufficient shielding effect, the detection efficiency for charged particles of one HVT is required to be better than 95%. All HVT detectors were produced by Beijing Hamamatsu photon techniques Incorporated.

Signals from HVT will be processed by HEB. When the amplitude of an HVT signal is higher than a preset threshold (adjustable by tele-commands), HEB will generate a veto signal with a preset width (also adjustable by tele-commands). Those good events coincident with the veto signal will be tagged as anti-coincidence events. These events will be rejected according to some certain rules during the ground-based data analysis.

**2.4.1 The top HVT**

Each top HVT is just above three HEDs and their collimators. Thus, six top HVTs fully cover the 18 HEDs and collimators, which can efficiently reduce the background caused by incident charged particles from the front. The distribution of top HVTs on HE is shown in Figure 1.

The geometric area of each top HVT is about 2250 cm$^2$ (roughly a fan-shaped plate). A 2-inch PMT (Hamamatsu R6231-07) is coupled in the middle of the HVT. Its structure is shown in Figure 8. Each top HVT adopts a BC-408 plastic scintillator produced by Saint-Gobain as the detector with a thickness of 6.4 mm. The upper and lower shells are aluminum sheets with a thickness of 1 mm. To lower the loss of transmissivity of X-ray photons, the aluminum sheet right above the collimator is removed and a polyethylene layer which is a protective coating for the plastic scintillator is exposed. The thickness of a plastic scintillator is also reduced to 3.4 mm in those areas.

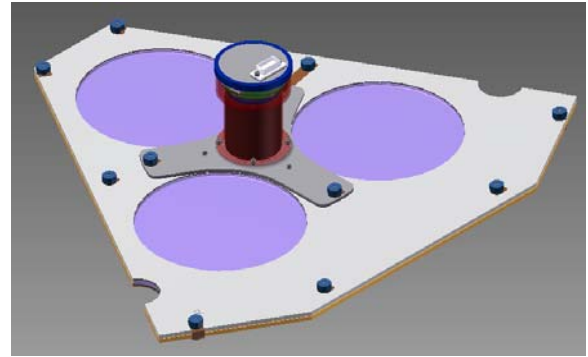

Figure 8. A sketch of a top HVT

**2.4.2 The lateral HVT**

There are 12 lateral HVTs closely arranged one by one. As shown in Figure 1, they fully cover the lateral sides of HEDs to efficiently reduce the HED background caused by high energy charged particles coming from lateral sides.

A lateral HVT is designed as a flat structure with an area of 1000 cm$^2$. A 2-inch PMT (Hamamatsu R6231-07) is coupled to read out signals. A lateral HVT structure is shown in Figure 9. A lateral HVT adopts the BC-408 plastic scintillator produced by Saint-Gobain. Its thickness is 6.4 mm and it is covered by a Ta shell with a thickness of 1 mm.

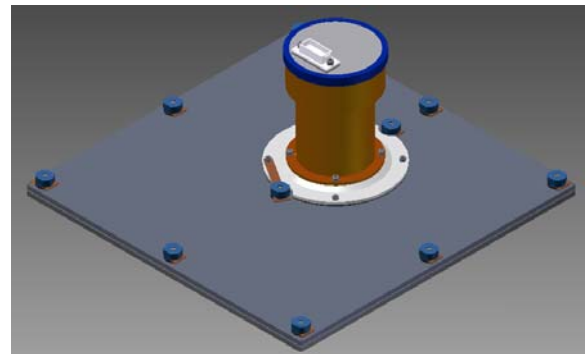

Figure 9. A sketch of a lateral HVT

**2.5 Particle monitor (HPM)**

The HE is configured with three identical HPMs to provide warning information about an abnormally increasing flux of high energy charged particles for the satellite platform. When the satellite is in orbit, especially passing the South Atlantic Anomaly (SAA) region, if the count rate of HPM exceeds the warning threshold (adjustable by tele-commands) in three consecutive seconds, the On-Board Data Handling (OBDH) system of the satellite platform will inform the HEB to switch off the high voltages of HEDs and HVTs to avoid damage.

An HPM consists of a small BC-440M plastic scintillator produced by Saint-Gobain, R647 PMT produced by Hamamatsu, front-end circuit, and other auxiliary structures. The plastic scintillator of HPM is machined into a cylinder with both diameter and height of 10 mm, see Figure 10. Except for the coupling surface with PMT, the plastic scintillator is painted with reflective paint and coated by Teflon as reflector. According to simulation, HPM is sensitive to electrons with energy higher than 1.5 MeV and protons with energy higher than 20 MeV.

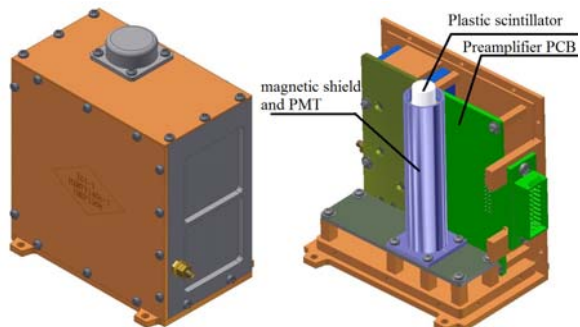

Figure 10. Left: the overall appearance of HPM; right: the inner components of HPM.

**2.6 Electronics subsystem**

The electronics subsystem is the control and information center of HE. It consists of a power-supply box (named HEA) and a signal procession box (named HEB). HEA provides power supplies for HEB and all detectors, executes remote control instructions, and generates part of telemetry signals.

HEB is the core of the HE signal processing. Taking into account the function and the requirement of reliability, it is divided into five units, including three Physical Data Acquisition Units (PDAU), one Anti-coincidence Shield Unit (ASU), and one Status Data Acquisition Unit (SDAU). There is a cold backup for each unit, respectively, to increase the reliability of the whole system, which can be switched on/off by uploaded telecommands.

SDAU is responsible for dealing with the pulse signals from three HPMs. The threshold of input signal can be adjusted by telecommands. If the signal amplitude of HPM exceeds the preset threshold, the corresponding counter will be added by one. SDAU collects the count rates of three HPMs every second and passes this information to the OBDH by the 1553B bus, which is one of the criteria used to decide whether to switch off the HVs of HEDs and HVTs.

ASU is responsible for dealing with the signals from the 18 HVTs and can independently communicate with OBDH by the 1553B bus. It is in charge of adjusting the high voltages, signal thresholds and veto signal widths of the 18 HVTs by telecommands. A signal of HVT exceeding threshold will produce a veto signal with the preset width and be delivered to PDAU. If there is a simultaneous HED signal, an anti-coincidence label will be set to the corresponding position of the event data packet. The count rates of the 18 HVTs will be recorded every second and passed to OBDH by the 1553B bus.

The three PDAUs are the heart of HEB. Eighteen HEDs and HGCs are respectively divided into three groups in the same way of collimators. Each PDAU is devoted to processing the signals of one group, which includes six HEDs and six HGCs. The HED signals exceeding the preset threshold (adjustable via uploaded commands) will be passed to PSD. If a signal's width is less than the preset PSD threshold, the event will be regarded as a good event and then be converted to 8-bit energy information by an ADC. Otherwise, it will be regarded as a CsI-involved event and be discarded. PDAU packs the information of a good event to form an 8-byte data packet. The data packets are arranged event by event to form a data flow. Finally, the data flow will be delivered to OBDH by the LVDS bus. Every "good event" data packet includes the following information:

a. Energy information：8 bits
b. Pulse width information：8 bits
c. Signal Tags of the 18 HVTs at the moment of this "good event"：18 bits
d. Channel information：5 bits
e. Information on event type ：2 bits
f. Timing information (sub-second part) ：19 bits
g. Cyclic Redundancy Check (CRC) code：4 bits

The timing information of HEB includes a second part and sub-second part. The sub-second part is generated by a 500 kHz clock from the frequency-division of a highly accurate 5 MHz clock from the satellite platform (in its working temperature -20°~ 45°, its short-term stability is $10^{-10}/20$ min and its long-term stability is $5 \times 10^{-9}$/day). Thus, there is 19-bit sub-second information in the data packet for each science event. The second information is provided by the PPS signal coming from the GPS. This is inserted into the science data flow as a special event. The accurate arriving time of each science event is determined by the sub-second information plus the second information from the previous and nearest GPS second event. The timing accuracy of science events relative to GPS is determined by the frequency of the sub-second clock, which is ±2 μs for HE.

The satellite platform broadcasts UTC time every second via a 1553B bus to synchronize all the payloads. After receiving the synchronization command, HEB records the UTC time and the current GPS second information into engineering data packet and delivers to OBDH via the 1553B bus.

**2.7 Operation mode**

In the application of NaI(Tl)/CsI(Na) phoswich, CsI(Na) is usually used as an active shielding to depress the Compton effect by Pulse Shape Discrimination (PSD). By setting the PSD threshold to a larger value via tele-commands, the CsI(Na) events will be changed to "good events". Then all the events from both NaI(Tl) and CsI(Na) will be recorded and the data packets will be downloaded completely. This feature can be used to detect GRBs, benefiting from the large effective area of the CsI(Na) crystal to MeV gamma-rays. The thickness of CsI(Na) is 40 mm and its total geometric area is about 5100 cm$^2$. The spectra of GRB prompt emission mainly spans from ~keV to ~MeV. According to simulations, HED CsI(Na) has excellent ability in timing, spectra and localization of GRBs in about 0.2-3 MeV, which will enhance the scientific returns of HE.

In normal mode of HE, the measured energy range of CsI(Na) is about 40-600 keV (note that this is deposited energy). Photons with energy less than 200 keV are very unlikely to penetrate the structure of HXMT and impinge on CsI(Na) detectors. Given that the peak energies ($E_{peak}$) of the majority of GRB spectra fall in the range of about tens keV to ~MeV [12,13], in order to obtain a high-quality GRB spectrum, the CsI(Na) energy band should cover the range from several tens of keV to several MeV. Therefore, a dedicated mode for HE is designed to optimize the spectral measurement. To achieve this goal, the high voltage of HEDs are reduced to achieve a gain of one fifth of that in normal mode and disable the auto-gain control at the same time via tele-command; thus this mode is called Low Gain mode (previously called GRB mode). As a result, there are two operation modes for HE: normal mode and low gain mode, listed in Table 2. NaI(Tl) will lose its coverage in the low energy band (20-100 keV) in low gain mode, thus NaI(Tl) will become less effective than CsI(Na) in low gain mode. By contrast, CsI(Na) is effective for monitoring the gamma-ray sky in both normal mode and low gain mode.

Table 2. Working modes of HE telescope

|  | Normal mode | Low Gain mode |
| --- | --- | --- |
| NaI Energy Range | 20-250 keV | 100 – 1250 keV |
| CsI Energy Range | 40-600 keV | 0.2 – 3 MeV |
| HV of PMT | normal | reduced |
| Auto-gain control | Enabled | Disabled |

Note: The above energy ranges are approximate; the exact value varies among the 18 detectors.

## 3 On-ground calibration

Before the launch, the flight modules of HE have passed thermal and vibration tests, together with satellite-system-level integration ones. The performance of HE is confirmed by on-ground calibration experiments.

The determination of energy and timing responses is the main purpose of the HE on-ground calibration. The calibration experiments have been carried out both in normal mode and low gain mode. The experiments in normal mode were implemented with a double-crystal monochromator and several radioactive sources listed in Table 3. Only radioactive sources were used in the calibration experiments in low gain mode since it is very hard to obtain a monochromatic X-ray beam with high enough energy via a double-crystal monochromator. During calibrating the normal mode, the relative position between HGC and HED was fixed to be the same status as inflight. AGC was kept on to ensure the HED gain did not change during all the calibration campaign, performed at the National Institute of Metrology in Beijing of China. The laboratory temperature was controlled at 18±2℃, the same of the working one of HED in orbit.

### 3.1 Calibration facility

#### 3.1.1 Double-crystal monochromator

To accurately calibrate the energy response of HED, especially for energy around the Iodine K-edge (~33.18 keV), a dedicated double-crystal monochromator based on an X-ray tube was built. Its basic principle is shown in Figure 11[14, 15].

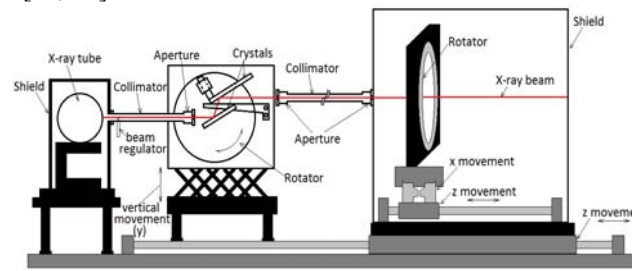

Figure 11. A sketch of the double-crystal monochromator.

Si-111 and Si-220 crystals have been used to produce a continuously adjustable monochromatic X-ray beam in the range 15-160 keV. The monochromaticity is 0.6% at 60 keV for this facility [15]. The diameter of the X-ray beam size used in experiments is 6 mm. The calibration experiments were performed at 25 energy points in the 15-160 keV energy band. The interval between energy points is ~1 keV around the Iodine K-edge.

For the calibration at each energy point, the X-ray beam was sequentially injected on 27 points sampled on the HED surface by means of computer-controlled movements. The accumulated spectrum from these 27 points was considered as the response of the whole HED plane to a parallel X-ray beam uniformly illuminating the surface of HED. The distribution of these 27 points is shown in Figure 12.

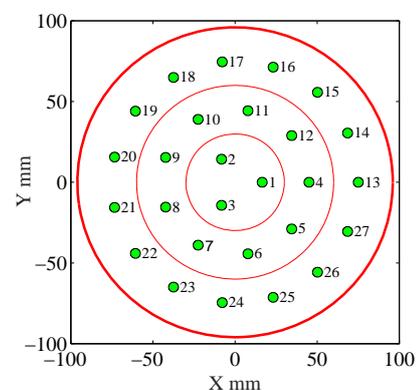

Figure 12. Position distribution of incident X-ray beam.

During the calibration campaign we found that the HED background did not change along with the Bragg angles under the same voltage and current of the X-ray tube. Therefore, at every voltage and current of the X-ray tube, we only measured the background once.

After changing the voltage of the X-ray tube, the actual spectrum of the X-ray beam was measured by a standard detector (HPGe detector of GL0110P model, produced by Canberra industries Incorporated). The HPGe detector was carefully calibrated with $^{241}$Am、$^{57}$Co、$^{152}$Eu、$^{109}$Cd, and $^{133}$Ba before the HED calibration experiments. This standard detector was the baseline of the energy response and detection efficiency of the X-ray beam for HED calibration.

### 3.1.2 Radioactive sources

The radioactive sources used in the calibration experiments were located in ampoule bottles in solution form. The uncertainty of radioactivity is less than 3%. The radioactive sources and HED were installed on special wooden holders. In normal mode, the distance from the center of the radioactive source to the surface of the HED was 1.5 meters; in low gain mode, this distance was 1.0 meter. The radioactive sources used in the on-ground calibration experiments are listed in Table 3.

Table 3. Radioactive sources used during on-ground calibration campaign

|  | Nuclide | Half-life period | Energy (keV) | Radioactivity (Bq) |
|---|---|---|---|---|
| Normal mode | $^{139}$Ce | 137.6 d | 165.9 | 3.004*10$^5$ |
|  | $^{133}$Ba | 10.5 yr | 81 356 | 1.188*10$^6$ |
|  | $^{57}$Co | 271.7 d | 122 | 8.699*10$^5$ |
|  | $^{241}$Am | 432 yr | 59.5 | 8.8089*10$^5$ |
|  | $^{109}$Cd | 461 d | 88 | 6.184*10$^5$ |
| Low Gain mode | $^{137}$Cs | 30 yr | 662 | 2.233*10$^6$ |
|  | $^{113}$Sn | 115 d | 392 | 1.519*10$^6$ |
|  | $^{60}$Co | 1925 d | 1173 1332 | 4.729*10$^5$ |
|  | $^{88}$Y | 107 d | 898 1836 | 7.697*10$^4$ |
|  | $^{40}$K | 1.28B yr | 1461 | (from environment) |

## 3.2 Calibration results and instrument performance

### 3.2.1 Pulse shape discrimination

Some incident X-ray photons deposit their energies only in the NaI(Tl) detector of HED, which are called NaI events or "good events"; some deposit their energies only in the CsI(Na), which are called CsI events; some deposit energies in both crystals, which are called Compton events or mixed events. NaI events are the science events to be recorded, while CsI events and mixed events are treated as shielding events. Due to the different decay time constants of NaI(Tl) and CsI(Na) scintillators, PSD can efficiently distinguish different types of events according to the pulse width. PSD adopted in HEB is so called the front-rear-edge discrimination technique [16]. Selecting a point respectively at the front edge and rear edge of a HED signal pulse, for which the amplitudes relative to the peak value of the signal pulse are exactly equal to a preset ratio, the time interval between these two points is defined as the pulse width of this signal. Theoretically, the pulse width does not change with the energy of an incident X-ray photon and thus can efficiently determine whether the current event is a good event [16].

Figure 13 is a typical pulse width spectrum of a $^{133}$Ba radioactive source in the calibration experiment in normal mode. Figure 14 is a typical pulse width spectrum measured with a $^{137}$Cs source in the calibration experiment in low gain mode. Figure 15 shows the peak values of NaI(Tl) and CsI(Na) events in the pulse width spectrum as functions of energy given by background data in low gain mode. The error bars stand for 2*FWHM of the pulse width of these two types of events. The events with the pulse width of 54-70 ch are NaI events and those with the pulse width of 90-120 ch are CsI events, roughly. The events between them are mixed events. The peak values of the pulse width of these two events are stable and do not change with the high voltage of HED and the energy of incident photons, which is consistent with the expectation.

In the pulse width spectrum, the peak value of NaI events is $PW_{NaI}$, and the width is $FWHM_{NaI}$. The peak value of CsI events is $PW_{CsI}$, with the width of $FWHM_{CsI}$. The figure of merit of PSD is defined as Equation 1.

$$A_{PHA} = \frac{PW_{CsI} - PW_{NaI}}{FWHM_{CsI} + FWHM_{NaI}} \qquad (1)$$

According to the data analysis of the system integration test, the $A_{PHA}$ values of 18 HEDs are distributed between 2.6 and 3.0, which indicates the high consistency of all HEDs and the good performance of the PSDs.

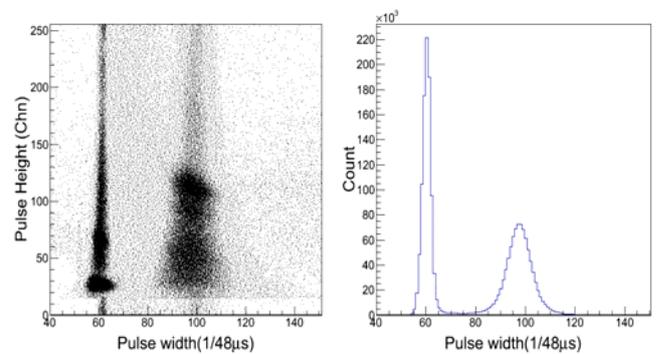

Figure 13. Pulse width spectrum of $^{133}$Ba in normal mode

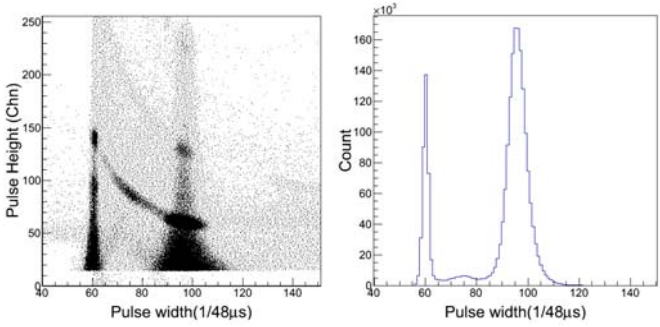

Figure 14. Pulse width spectrum of $^{137}$Cs in low gain mode

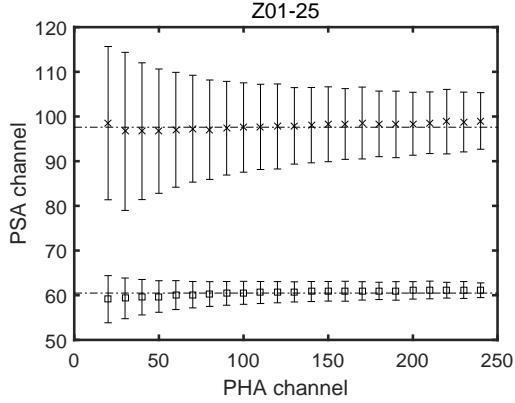

Figure 15. Position of the pulse width peaks of NaI(Na) and CsI(Na) events as functions of PHA derived from background data. The error bars stand for the 2*FWHM of the pulse width of these two types of events.

### 3.2.2 Energy-Channel (E-C) relationship

Many previous studies have shown the existence of non-proportionality between the scintillation efficiency of an NaI(Tl) crystal and the deposited energy of X/γ ray, especially for the energy around the Iodine K-edge [5, 6, 17-21]. Figure 16 presents experimental results of non-proportional scintillation response of NaI(Tl). The vertical axis stands for the pulse height per unit energy and all points are normalized to the value at 60 keV. It can be found that the result of HE is consistent with other experiments, except for the low energy band of Fermi/GBM.

According to the fits of the full-energy-peak of the NaI events of HED, we can obtain the HED response to different X-ray energies and the E-C relationship.

To fit the E-C relationship in normal mode, we adopted a piecewise fitting method. According to the non-proportional efficiency curve in Figure 16, there are three segments in the observable energy range of HED, i.e., E < 33.17 keV, 33.17 keV ≤ E < 50.2 keV, and E ≥ 50.2 keV. Each segment is fitted by a second-order polynomial as shown in Equation (2). In low gain mode, the E-C relationship is fitted by a linear function.

$$E(ch) = p_1 * ch^2 + p_2 * ch + p_3 \qquad (2)$$

The E-C relationship of detector Z01-25 is shown in Figure 17. The residual of the fit is smaller than 1% at all energy points. The E-C relationships of all other HEDs are dealt with using a similar method.

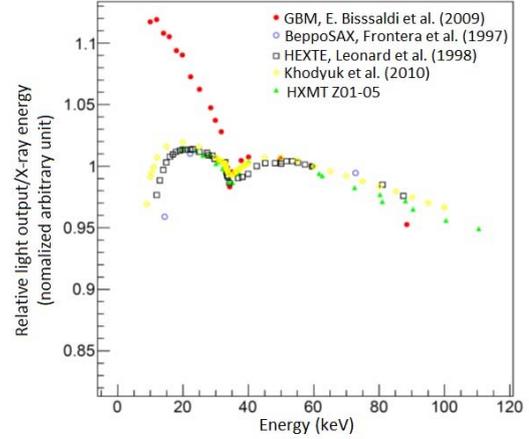

Figure 16. Nonlinearity response of NaI(Tl) for HED Z01-25, normalized to unity at 60keV. It is consistent with some of previous studies, except for the low energy band of GBM.

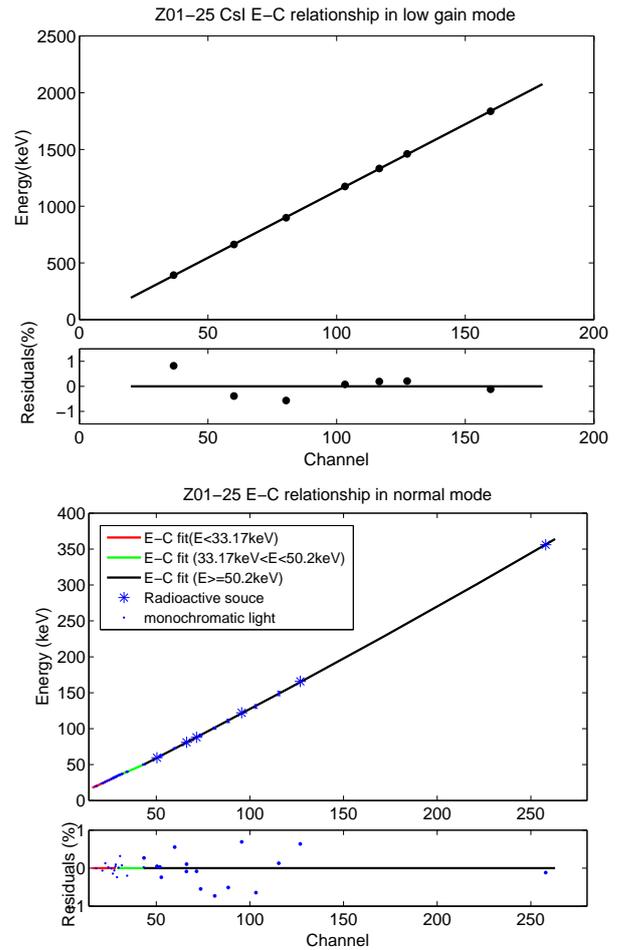

Figure 17. E-C relationship and the corresponding residuals of detector Z01-25. Top panel: low gain mode; Bottom panel: normal mode.

According to the calibration results, the effective detection energy range of NaI(Tl) is about 16 - 350 keV for all the 18 HEDs in normal mode. For CsI(Na), it is about 130 keV - 3 MeV in low gain mode.

**3.2.3 Energy resolution**

Due to statistical fluctuation from the scintillation light yield and light collection, the monochromatic X-ray photons generate a peaked spectrum with a certain width in HED, which can be normally fitted by a Gaussian function. For a linear system, the intrinsic energy resolution $\eta$ of a detector is a power-law function of energy with an index of -0.5 [6, 11]. However, the energy resolution of a realistic detection system will be impacted by several factors and deviates from this law [5, 6].

In normal mode, the change of the energy resolution is discontinuous around the Iodine K-edge. Thus, the relationship of the energy resolution versus energy was fitted separately below and above the Iodine K-edge. The adopted fitting function is provided in Equation 3,

$$\eta(E) = \frac{\sqrt{a + bE + cE^2}}{E}. \quad (3)$$

For the calculation of the energy resolution, the double-crystal monochromator beam dispersion has to be taken into account. In principle, if $\omega_1$ is the FWHM width of the full-energy-peak measured by HED, and $\omega_2$ is the FWHM width of the X-ray beam measured by the HPGe standard detector, then the intrinsic energy spread of HED at energy $E$ is indicated by $\omega_3^2 = \omega_1^2 - \omega_2^2$.

In low gain mode, the energy of incident photons is relatively high. The adopted fitting function of $\eta(E)$ is as written in Equation 4,

$$\eta(E) = a * E^b. \quad (4)$$

The upper panel of Figure 18 is the energy resolution curve of detector Z01-25 in normal mode. The lower panel is the energy resolution (as a function of E) measured by radioactive sources in low gain mode. The energy resolution can be defined as FWHM in ADC channel divided by the peak centroid channel. It can also be defined as FWHM in keV divided by the energy of the incident X-ray photons. In this paper, we use the former definition.

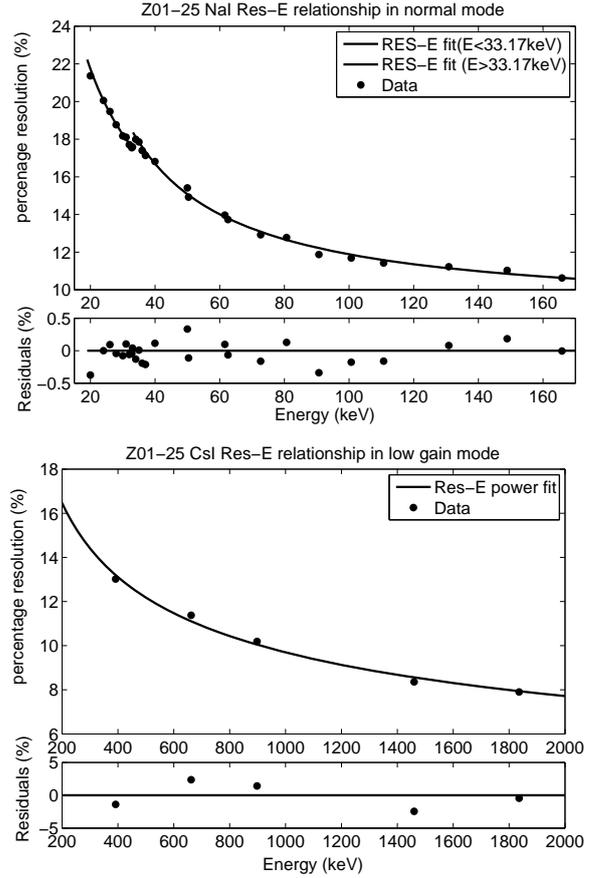

Figure 18. Energy resolution of detector Z01-25 as a function of energy, Top panel: normal mode; Bottom panel: low gain mode.

**3.2.4 Detection efficiency**

The detection efficiency of HED in normal mode was calibrated with the double-crystal monochromator. The intensity of the monochromatic light is described as in Equation 5,

$$I(E_i) = \frac{N_0(E_i)}{eff_0(E_i)}, \quad (5)$$

where $N_0(E_i)$ is the count rate of the full-energy-peak (average energy is $E_i$) measured by the HPGe standard detector, and $eff_0(E_i)$ is the detection efficiency of HPGe to the full-energy-peak of the X-ray photons at energy $E_i$. The detection efficiency of HPGe is referenced from Liu 2016 [22]. As shown in Figure 19, the black circles are the data of radioactive sources, which are consistent with the result of Monte-Carlo simulations (the solid line). The dashed line stands for the total efficiency which includes all the X-ray events, not only those depositing all their energy in the detector, but also those depositing part of their energy in the detector.

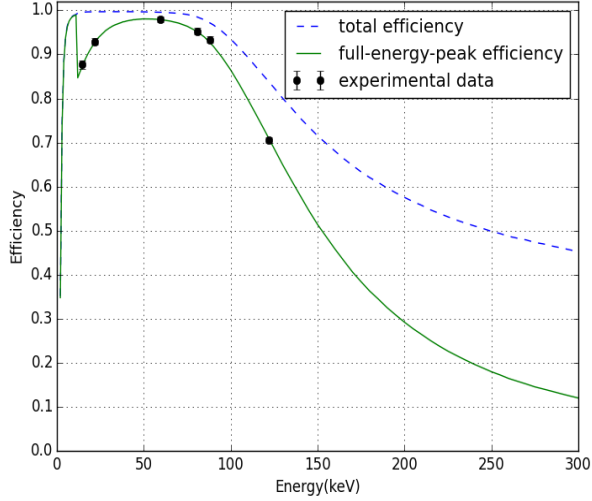

Figure 19. Efficiency of HPGe. Dashed line: the total efficiency curve from simulation; solid line: the full-energy-peak efficiency curve from simulation; solid circles: the full-energy-peak efficiency data from experiments.

Then the HED detection efficiency of full-energy-peak of the X-ray at energy $E_i$ is as shown in Equation 6,

$$eff_{HED}(E_i) = \frac{N_{HED}(E_i)}{I(E_i) * K_I}, \quad (6)$$

where $N_{HED}(E_i)$ is the count rate of the area of the full-energy-peak of HED calibration spectra (dead time corrected), and $K_I$ is the stability factor of the beam flux and is set to 1 because the beam flux did not show significant change in the calibration process.

In low gain mode, all calibration experiments adopted radioactive sources. The detection efficiency of CsI(Na) is expressed as

$$eff_{HED}(E_i) = \frac{N_{HED}(E_i)}{A_0 p(E_i) \omega}, \quad (7)$$

where $N_{HED}(E_i)$ is the count rate of the full-energy-peak of γ-ray with energy $E_i$, $A_0$ is the measured radioactivity (corrected by half-life period), $p(E_i)$ stands for the branching ratio of γ-ray with energy $E_i$ of a radioactive source, and $\omega$ represents the geometric factor of a radioactive source relative to the detector $\omega = \frac{1}{2}\left(1 - \frac{l}{\sqrt{r^2 + l^2}}\right)$, where $l$ means the distance from the center of a radioactive source to the surface of HED and $r$ stands for the radius of HED [22].

In normal mode, NaI events are good events. In low gain mode, CsI events are also good events. The top panel of Figure 20 shows the full-energy-peak detection efficiencies of 18 HEDs in normal mode. The red curve is the result of Monte-Carlo simulation, while the blue lines are the experimental result of 18 flight HED modules, which are very consistent with each other and consistent with the trend of the simulation result. The bottom panel of Figure 20 is the full-energy-peak detection efficiency of CsI(Na) in low gain mode. The red curve is the result of Monte-Carlo simulations, while the blue lines are the experimental results of the 18 flight HED modules, the trend of which is generally consistent with the simulation but the measured efficiencies are lower. The efficiency discrepancy is probably due to the signal selection process. To avoid wrong screening, signals with the width of 91~110 PSA-channel were considered as pure CsI events during the data analysis. As shown in Figure 15, the lower energy photons induced a broader distribution of pulse width and thus more low energy photons are excluded. As a result, the difference between the simulation and experimental results decreases with increasing energy.

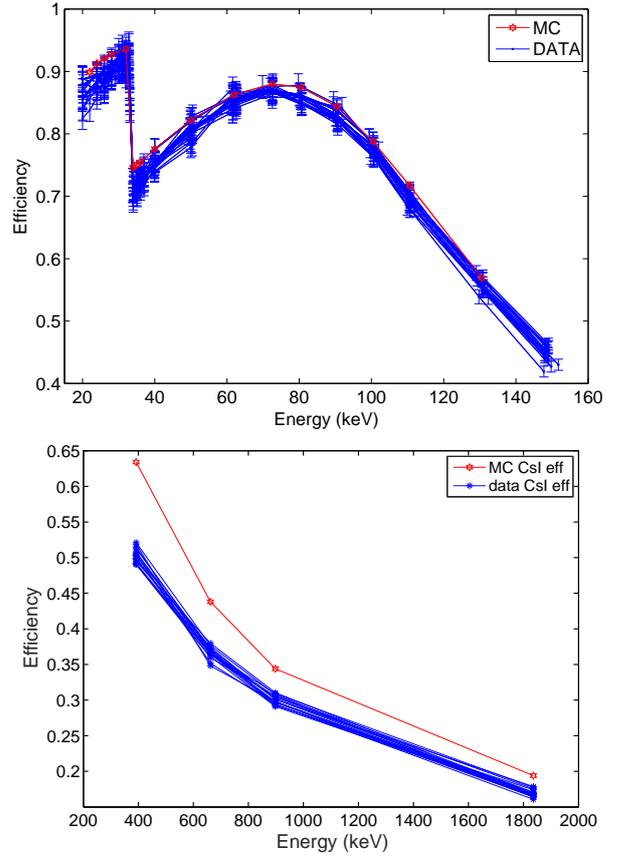

Figure 20. Detection efficiencies of 18 HEDs derived from calibration experiments; the red curve is from simulation. Top: normal mode; Bottom: low gain mode.

To quantify the impact of the top HVTs on the efficiency of HED, an HVT detector was installed in front of HED Z01-18 according to the configuration pattern on the satellite. Then the calibration experiments were repeated in the energy range of 28-150 keV to obtain the impact of HVT on the detection efficiency of HED. The results are shown in Figure 21. The simulation results with HVT are shown in the same figure and the efficiency is slightly higher than the experimental results. HVT indeed reduced slightly the full-energy-peak detection efficiency of HED. The detection efficiency in the low energy band decreases because of the loss of X-ray transmissivity, and because of the increase of Compton events in the high energy band.

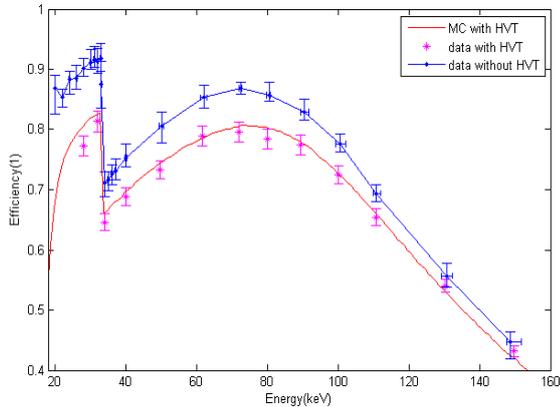

Figure 21. Comparison of efficiency between HED with and without HVT.

### 3.2.5 Response uniformity

To understand the overall performance and the non-uniform response of HED, 192 sampled points on the HED detector plane were sequentially irradiated with the X-ray beam at 50 keV. Typical results of HED detector Z01-5 are shown in Figure 22 to Figure 24. Figure 22 displays the response non-uniformity of the full-energy-peak, Figure 23 for the non-uniformity of energy resolution, and Figure 24 for the non-uniformity of detection efficiency. The experiment results show that the relative standard deviation of the full energy peak, energy resolution, and detection efficiency are smaller than 2%, 2.5%, and 2.5%, respectively. These results show that the HED scintillators contain no significant dead area, and the scintillation light is collected well.

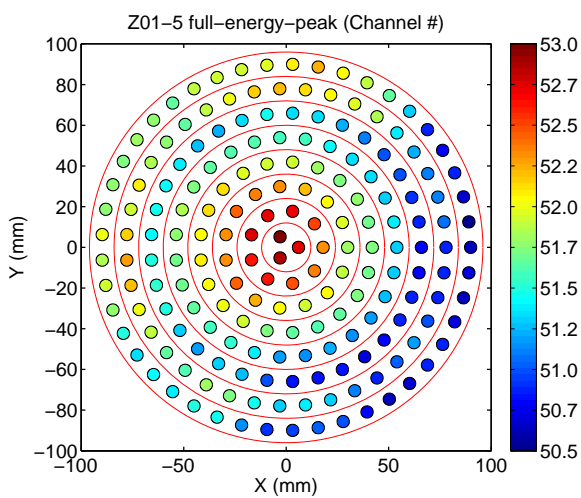

Figure 22. Peak mapping of 50 keV photons on 192 positions for HED Z01-5. The value of a color bar indicates the signal amplitude of 50keV photon in ADC channel.

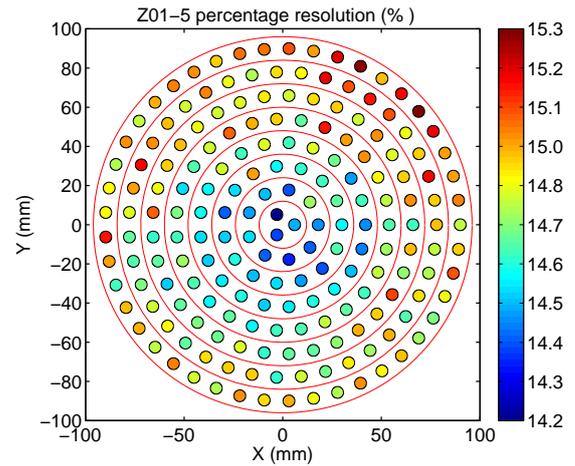

Figure 23. Energy resolution mapping at 50keV photons on 192 positions for HED Z01-5. The value of color bar indicates the energy resolution in terms of percentage.

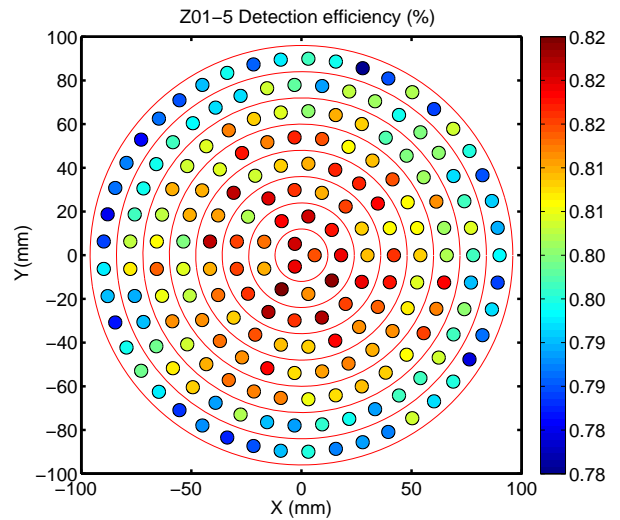

Figure 24. Efficiency mapping at 50 keV photons on 192 positions for HED Z01-5. The value of color bar indicates the detection efficiency in terms of percentage.

### 3.2.6 Timing response

#### 3.2.6.1 Timing accuracy

The accuracy of the arrival time of the events recorded by HE is determined by the sub-second module. This module adopted the 500 kHz clock obtained from the frequency-divided 5 MHz clock provided by the satellite platform. Thus, the intrinsic time resolution of the system should be ±2 μs. The timing information will be recorded in the data packet of every event and then be downlinked. During the integration test of the satellite, the timing accuracy was calibrated according to the interval of the PPS signals of GPS.

HE inserts the PPS signal of GPS into the science data flow as a type of special event and simultaneously records the sub-second information when GPS second arrives. The-oretically, the value of the sub-second counter between two adjacent GPS second events is always 500,000, since we use a 500 kHz clock to record the PPS. The counter deviation from 500,000 is considered as an indication of the system timing accuracy. Figure 25 is a distribution of the counter deviation recorded during an experiment of the satellite integration test, where the horizontal axis is the corresponding time fluctuation and the vertical axis is the histogram number. The timing accuracy of HE is ±2 μs relative to the GPS time, which is consistent with the theoretical expectation and satisfies the design requirement (better than 25 μs) of the timing accuracy.

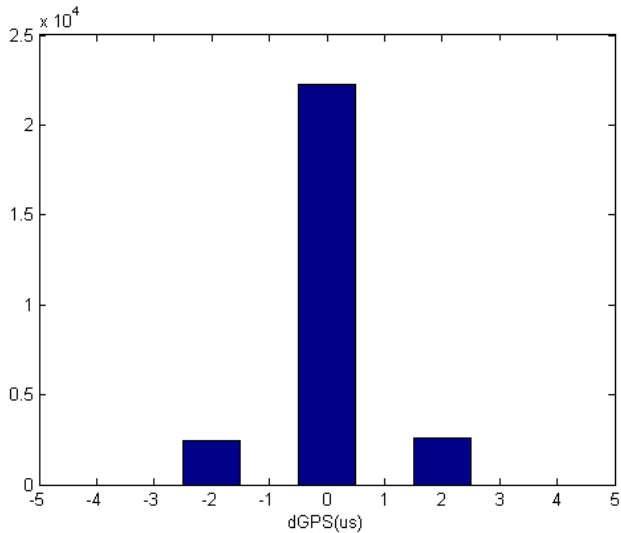

Figure 25. Histogram of intervals (the theoretical value subtracted) between two consecutive GPS events.

### 3.2.6.2 Dead-time

For each event, HEB needs some time to process the signal of HED. The simultaneous signals produced by the HEDs in the same group (see above) during this event processing will be discarded. Thus, this causes dead time of this HED. The dead time of one event is normally about 4~8 μs, but it will be longer for the signals generated by high energy charged particles. There is a deadtime counter in HEB for each HED. When the system is too busy to deal with the signal of a channel, the deadtime counter of this channel will count with the step of 1/8 μs to record the dead time. The total deadtime of a second will be finally recorded in the engineering data packet.

In principle, the X-ray counts detected by HED follow a Poisson distribution. It can be demonstrated that the time interval between two adjacent events follows a negative exponential distribution which indicates that the probability is high for the events with relatively shorter time intervals [11]. In the spectrum of the time interval of two adjacent events, the point deviating from a negative exponential distribution indicates the maximum dead time of this system to process a normal signal. Figure 26 gives the result with a set of background testing data. According to this result, the maximum dead time of HE to process a normal NaI(Tl) event is smaller than 8 μs.

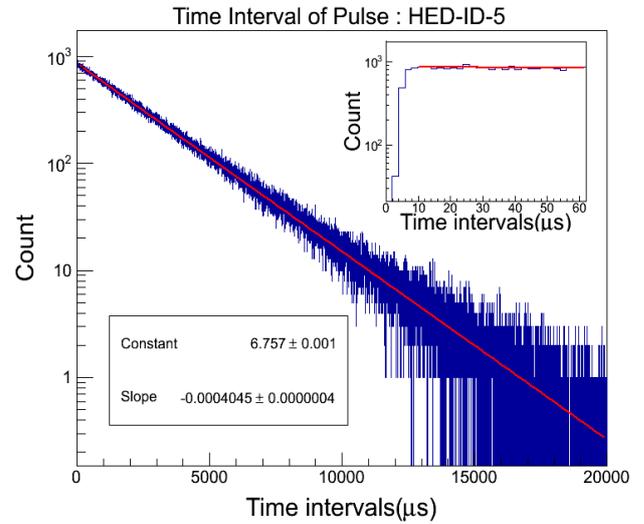

Figure 26. Statistics of the time interval between adjacent photons. It is well fitted with a line as expected in a logarithmic coordinate system.

## 4 Conclusions

Before the launch on June 15, 2017, manufacturing and environment tests of HE flight-modules were performed. Now all the HE instruments work properly in space. The energy response, timing response and detection efficiency of HED are measured in a series of on-ground calibrations. In these calibration tests, the double-crystal monochromator is mainly used to cover the 15-160 keV band and radioactive sources have been used to produce more complementary energy points. The analysis shows that the instrument satisfies the design requirements.

In normal mode, the energy band of HE is about 16-350 keV according to the results of calibration. When the target source is obscured by the Earth in a pointed observation or when normal mode of HE is not very useful in some observations, HE will be configured into low gain mode by tele-commands. Either in normal mode or low gain mode, HE serves as a GRB monitor with a large FOV and large area. The mean effective area is about 1000 $cm^2$ for photons in ~ 200-3000 keV. Especially for energies higher than 1 MeV, it is significantly more sensitive than Fermi/GBM and Suzaku/HXD (WAM), therefore it can play an important role in the observation of GRBs and the electromagnetic counterparts of gravitational wave events.

# Acknowledgement

*The authors express thanks to the people helping with this work, and acknowledge the valuable suggestions from the peer reviewers. This work was supported by China National Space Administration (CNSA) and the Chinese Academy of Sciences (CAS).*